\documentclass[twocolumn,pra,showpacs]{revtex4}
\usepackage{graphicx}

\begin{document}

\title{Controlled splitting of an atomic wave packet}
\author{M. Zhang,$^1$ P. Zhang,$^{2,3}$ M. S. Chapman,$^{3}$ and L. You$^{1,3}$}
\affiliation{$^1$Center for Advanced Study, Tsinghua University,
Beijing 100084, P. R. China}
\affiliation{$^2$Institute of Theoretical Physics, The Chinese
Academy of Sciences, Beijing 100080, P. R. China}
\affiliation{$^3$School of Physics, Georgia Institute of
Technology, Atlanta, GA 30332, USA}

\date{\today}

\begin{abstract}
We propose a simple scheme capable of adiabatically splitting an atomic
wave packet using two independent translating traps.
Implemented with optical dipole traps, our scheme allows a high degree of
flexibility for atom interferometry arrangements and highlights
its potential as an efficient and high fidelity atom optical beam splitter.
\end{abstract}

\pacs{03.75.Be, 39.20.+q, 39.25.+k, 03.65.-w}

\maketitle

Controlled manipulation of atomic motion is an active frontier in
{\it atom optics} \cite{berman97}, an emerging field built upon
the close analogy between matter waves and optical fields
\cite{pfau}. In recent years, the field of {\it atom optics} has
received a significant boost from atomic Bose-Einstein
condensates, the matter wave analogy of a laser field. The
unavoidable dispersion associated with a massive field has limited
the performance of various atom optical elements, such as atomic
mirrors and beam splitters. An innovative approach involves
trapped or guided atoms, as a recent experiment by H\"{a}nsel
\textit{et al.} has demonstrated, splitting and
uniting a trapped rubidium cloud in a chip trap
\cite{hansel01}.

\begin{figure}
\includegraphics[width=3.25in]{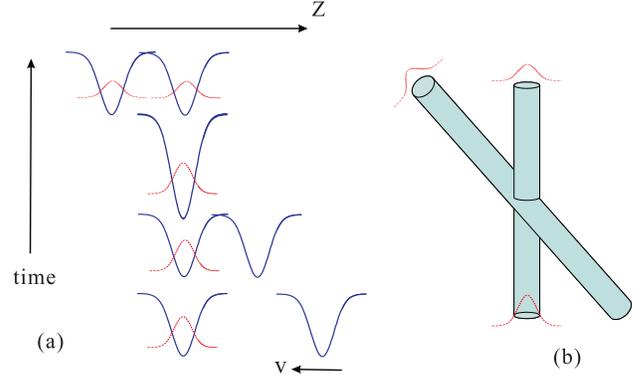}
\caption{(a) Illustration of the proposed splitting
scheme with two moving traps; this is in fact the familiar type
X-cross splitting for an guided atom as in (b).}
\label{fig1}
\end{figure}

Atomic beam guiding, or confined propagation, is a well studied
topic in atom optics. A variety of setups, involving a single
current carrying wire \cite{sw} to multiple wires \cite{hinds},
hollow optical fibers \cite{hole,holeE}, and far off-resonance
optical traps (FORT) \cite{fort,fort2} have been thoroughly
investigated. Several schemes for coherent large angle splitting
of a guided atomic wave packet have also been realized
experimentally \cite{muler00}. In this paper, we outline a simple
scheme capable of coherently splitting an atomic wave packet using
two identical translating traps. As illustrated in Fig.
\ref{fig1}, our system involves a trap with an atom and an
identical trap with no atom. By adiabatically translating the
empty trap towards the trap with the atom, overlapping them and
then passing over to the other side, the atom will be coherently
split into the two traps depending on the magnitude of the
relative velocity. Such a situation can be easily implemented
using optical dipole traps as in Ref. \cite{ketterle,ober} or
micro-traps in simple conveyor belt configurations
\cite{hansel01,bonn,Niu}. In addition to the above translating
trap system, our model also applies to situation (b) of Fig.
\ref{fig1} where an atomic wave packet propagates through an
X-cross beam splitter.

In the following, we carefully describe our
scheme. For simplicity, the system is taken to be
one dimensional along the z-axis, which is easily realizable
through tight confinement in the other two orthogonal directions.
In the center of massive motion frame, we
take the origin to be at middle point of the two traps
$V(z-d/2)$ and $V(z+d/2)$, centered respectively at $z\pm d/2$.
The two traps start at a large distance $d$ away. They
approach each other, overlap, then pass on, and move away from each other.
Because of the symmetric arrangement, parity is conserved at any time $t$,
thus can be used to characterize the eigen-structure of our system.
For large negative/positive values of $d$, when the trap at $z-d/2$
is at far left/right, the associated Hilbert space is rather simple
and consists of doubly degenerate states of the
single trap energy levels $\epsilon_n$. The corresponding
single trap eigen-state $\phi_n(z)$ satisfies
\begin{eqnarray}
\left[{p_z^2\over 2M}+V(z)\right]\phi_n(z)=\epsilon_n \phi_n(z).
\label{ss}
\end{eqnarray}
With both traps, $V_2(z,d)=V(z-{d/2})+V(z+{d/2})$, we find
\begin{eqnarray}
\left[{p_z^2\over 2M}+V_2(z,d)\right]
\psi_n^{(p)}(z,d)=E_n^{(p)}(d) \psi_n^{(p)}(z,d),
\label{sd}
\end{eqnarray}
with an eigen-energy $E_n^{(p)}(d)$ and an eigen-function $\psi_n^{(p)}(z,d)$
parametrically dependent on $d$.
The spectrum is $E_n=\epsilon_n$
for large separations $d$ when quantum tunnelling is practically absent.
The double degeneracy is due to the presence of both
an even/odd parity ($p=e,o$) state for each level with $\epsilon_n$, {\it i.e.},
\begin{eqnarray}
E_n^{(e/o)} (z,d)&\stackrel{|d|\to\infty}{\longrightarrow}& \epsilon_n,\nonumber\\
\psi_n^{(e/o)}(z,d) &\stackrel{|d|\to\infty}{\longrightarrow}& {1\over\sqrt 2}
\left[\phi_n(z+{d/2})\pm \phi_n(z-{d/2})\right].\hskip 24pt
\end{eqnarray}
When $d=0$, the
eigen-structure of Eq. (\ref{sd}) is simply that
of an atom in a single trap $2V(z)$
(or two completely overlapped single traps) and consists of
alternating even and odd parity states with increasing energies.
Since the parity is conserved for a symmetric single trap
$V(-z)=V(z)$, the $d$-dependent adiabatic energy level diagrams
is obtained simply by connecting
energy levels in the limiting cases
for $d=-\infty$, $0$, and $\infty$. In ascending order,
they are $E_0^{(e)}\le E_0^{(o)}<E_1^{(e)}\le E_1^{(o)}<E_2^{(e)}\cdots$.

\begin{figure}
\includegraphics[width=3.25in]{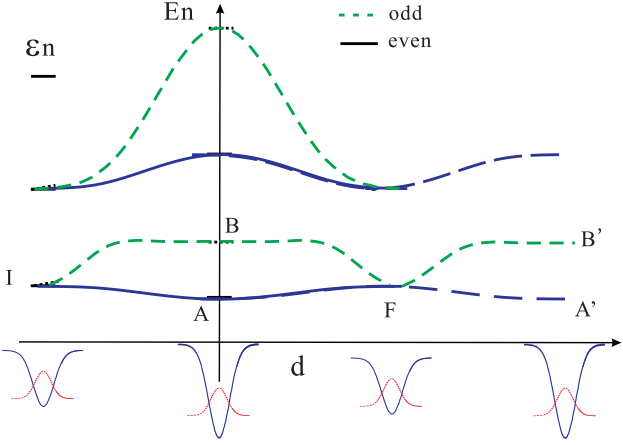}
\caption{The adiabatic energy levels in the center of mass frame.
The long dashed part to the right is a replica of negative $d$
on the left.
This figure should be compared to the analogous
Fig. 4 of Ref. \cite{hansel0102} and Fig. 1(c) of Ref. \cite{c2}.
}
\label{fig2}
\end{figure}

Our splitting scheme can now be easily presented.
Assuming an atom is initially in the ground motional state
$|\phi_0(z-d/2)\rangle$ of the left trap at a large negative value of $d$,
{\it e.g.}, at point $I$ in Fig. \ref{fig2}, its state is
simply in an equal superposition of states
$|\psi_0^{(e)}(z,d)\rangle$ and $|\psi_0^{(o)}(z,d)\rangle$, or
\begin{eqnarray}
\Psi(z,d)&\stackrel{d\to \rm -large}{\longrightarrow}&\phi_0(z-{d/2})\nonumber\\
&=&{1\over\sqrt 2}\left[\psi_0^{(e)}(z,d)+\psi_0^{(o)}(z,d)\right]. \hskip 12pt
\end{eqnarray}
When the right trap is adiabatically translated towards the left trap
and eventually passed over to the other side,
the two states of the superposition follow the adiabatic energy
levels $E_0^{(e)}$ or $E_0^{(o)}$, {\it i.e.}, along the
curves $I\to A\to F$ or $I\to B\to F$, respectively.
In the end when $d$ takes on a large positive value,
corresponding to the initial right trap now on the far left
side of the left trap, the atomic wave packet is split.
Depending on the relative phases of states
$|\psi_0^{(e)}(z,d)\rangle$ and $|\psi_0^{(0)}(z,d)\rangle$,
the superposition for atomic wave packets in the
two traps can be controlled. We find that
\begin{eqnarray}
\Psi(z,d)&\stackrel{d\to \rm +large}{\longrightarrow}&
{1\over\sqrt 2}\left[e^{-i\int E_0^{(e)}[d(\tau)]d\tau/\hbar}
\psi_0^{(e)}(z,d)\right.\nonumber\\
&&\hskip 16pt\left.+e^{-i\int E_0^{(o)}[d(\tau)]d\tau/\hbar}
\psi_0^{(o)}(z,d)\right], \hskip 24pt
\end{eqnarray}
where $E_n^{(p)}[d(\tau)]$ denotes the parametric time dependence
of eigen-energies on $d$ and the integration
is over the elapsed time. The splitting is
controlled by the relative phase between the two amplitudes; or,
\begin{eqnarray}
2\theta(t)={1\over \hbar}\int\left[E_0^{(o)}[d(\tau)]-E_0^{(e)}[d(\tau)]\right]d\tau.
\end{eqnarray}
For our scheme, this is simply
related to the area enclosed by the closed loop $I\to A\to F\to B\to I$; or with
$d{\cal A}$ as the surface integration measure,
\begin{eqnarray}
2\theta_C={1\over v\hbar}\oint_{I\to A\to F\to B\to I}d{\cal A},
\label{2t}
\end{eqnarray}
where a constant velocity $v$ is assumed.
Although not
exactly a topological invariant, it is important to note that
Eq. (\ref{2t}) does not depend sensitively on the details of the splitting process.
In the end, we find that the atomic wave packet is described by
\begin{eqnarray}
&&{1\over\sqrt 2}\left[e^{i\theta_C}
\psi_0^{(e)}(z,d)+e^{-i\theta_C}
\psi_0^{(o)}(z,d)\right]\nonumber\\
&\stackrel{d\to \rm +large}{\longrightarrow}&
\cos({\theta_C})
\phi_0(z+{d/2})
+i\sin({\theta_C}) \phi_0(z-{d/2}), \hskip 24pt\nonumber
\end{eqnarray}
apart from an overall phase. For
practical implementations, we note that the relative velocity
is only required to be a constant over a small distance,
where inter-well tunnelling coupling is non-negligible.

Our scheme can be contrasted with the works of Refs. \cite{hansel0102,c2}
where interferometry for a trapped atom were discussed based on
the splitting of a single trap into two separate traps.
In fact, the level diagram of negative $d$, replicated as the
long dashed lines to the right part of Fig. \ref{fig2}
was shown in Fig. 4 of Ref. \cite{hansel0102} and in Fig. 1(c) of Ref. \cite{c2}.
In the interference scheme proposed in Ref. \cite{hansel0102},
a ground state atom is split from a
splitting of the single trap (at point A) into two separate traps
(at point F) following adiabatic parametric motion along the d-axis.
A phase shift between the two separate traps
then appears as a finite population
in the first excited state (point B') upon recombing
into a single trap after moving back to A'.
The first half involves
atomic splitting, the time phase of the adiabatic eigenstate can be
neglected assuming the atom
starts and remains in the ground state. In the second half when
the double traps are reunited, the relative time phase
between the excited state (odd parity state along the F$\to$B' curve)
and the ground state (even parity state along the F$\to$A' curve)
may become significant. This will potentially
hinder an accurate determination of the small differential
phase accumulated (at point F) between the two parts of
the atomic wave packet in the two separate traps.
In the guided interference scheme of two connected
Y-shaped beam splitters \cite{c2},
the atom again starts at $d=0$ or
the trunk of the Y-shaped guide. It splits into
two wave packets along the two branches upon parametric propagation
 into the cross (regions of large $d$).
The second (inverted) Y-shapesd guide then reverses the
above and the two parts of the guided wave packets reunite
and interfere at $d=0$. Both works \cite{hansel0102,c2}
involve an open-ended parametric path for atomic splitting and
recombination and neither attempts an active utilization
of the relative time phase difference between different
adiabatic energy levels. In contrast, our scheme makes use
of a closed path in
the adiabatic energy level diagram (Fig. \ref{fig2}).
We start with two separate traps when the
lowest two adiabatic energy levels are degenerate.
The splitting is accomplished through the crossing of the
two traps and is controlled by the total time phase difference.
As a result, our scheme does not depend on the details of the exact
path, provided the parametric motion is adiabatic and
the parity of the system is conserved.

We now consider a realistic example easily implemented with two
optical traps. Assume each trap is
an inverted Gaussian formed by a red detuned focused laser beam,
{\it i.e.}, we take $V(z)=-M\omega^2\sigma^2e^{-z^2/(2\sigma^2)}$, which
is approximately harmonic near the bottom
$\sim M\omega^2z^2/2$ with a
ground state size $\sim \sqrt{\hbar/(M\omega)}$.
The width of the trap, $\sigma$, is another independent parameter.
For $d\gg\sigma$,
there are two symmetrically located minimums approximately
at $\pm d/2$; the merging of the two traps with decreasing
$d$ leads eventually to the merging of the two minimums
into just a single one located at the middle between the two
traps. This occurs at $d=\pm2\sigma$, as simply obtained
from ${d^2}V_2(z,d)/dz^2|_{z=0}=0$.

\begin{figure}
\includegraphics[width=3.25in]{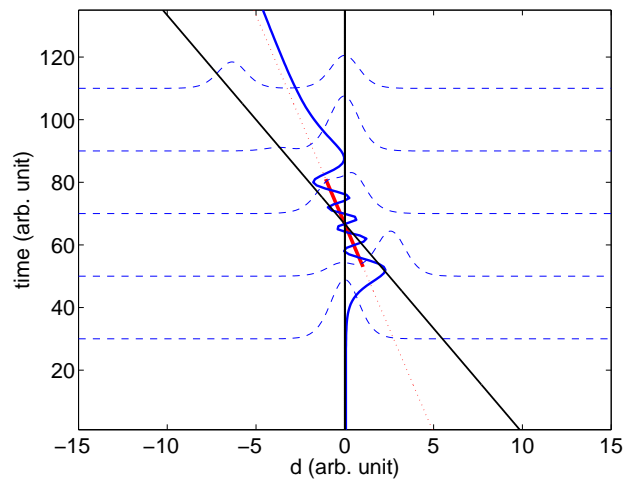}%
\caption{An illustration of the splitting in terms of the center of
mass coordinate of an atom originally in the left well.
}\label{fig3}
\end{figure}

Over a wide range of realistic parameters,
we find the splitting proceeds exactly as our scheme predicts,
provided adiabaticity is maintained.
In Figure \ref{fig3}, we illustrate the time dependent
splitting as obtained from numerically solving the time
dependent Schrodinger equation with translating traps.
The vertical and oblique solid straight lines denote the
centers of the two traps as in Fig. \ref{fig1}. The dotted oblique
straight line in between denotes the middle point between
the two trap centers and the thickened middle section
denotes the single minimum of the combined traps when $|d|\le 2\sigma$.
The thick wavy line denotes the center of mass motion
of an ground state atom initially in the left trap.
It does not follow the middle dashed line because of the
unbalanced splitting for the particular choice of $v=0.15$.
It displays clear tunnelling oscillations in the central region
when the two traps are close, especially where only a single
minimum exists. With increased speed $v$, the splitting
actually becomes easier to simulate and control as
the tunnelling oscillation disappears. To help visualize the splitting,
we have over-plotted the atomic wave-packet densities
in dashed lines at selected times. For a $^{87}$Rb atom,
this simulation corresponds to simply take $\sigma=\sqrt{\hbar/(M\omega)}$.
With $\omega=(2\pi)10$ (kHz),
we find $\sigma\approx 0.108$ ($\mu$m),
and the unit of time is $\approx 0.016$ (ms).
The trap speed of
$v=0.15$ corresponds to $\approx 1$ (mm/s).

\begin{figure}
\includegraphics[width=3.25in]{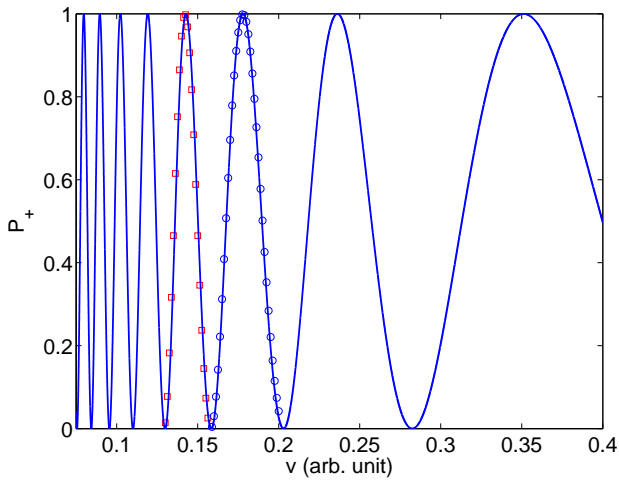}%
\caption{An illustration of our proposed atom splitter
where the probability of the atom to remain
in the stationary trap is plotted against the relative velocity.
}
\label{fig4}
\end{figure}

To have a desirable control over the splitting
requires $\theta_C\sim \pi$. This can be easily estimated from the total area of
${\cal A}\sim \omega\sigma$, which gives rise to a highest velocity value of $v\sim 1$,
or close to $1$ (cm/s), for the physical parameters considered above.
In Fig. \ref{fig4},
we display the predicted $1/v$ dependence of the relative phase Eq. (\ref{2t}).
The squares and circles denote results from
numerical simulations with 500 and 1001 spatial grid points.
Clearly, we see both convergence and agreement.
We have also attempted to optimize, a minimization of the
time required for the adiabatic splitting
using a more general time dependence for
$d[t]$, {\it e.g.}, with the use of Blackman pulses \cite{hansel01}
or other optimal control schemes \cite{zolleropt}. They did not
seem to significantly improve our result.

Intuitively, the surprising resistance of our scheme
to nonadiabatic transitions
can be simply summarized as follows:
1) parity conservation
forbids transitions between the two states
$|\psi_0^{(e)}\rangle$ and $|\psi_0^{(o)}\rangle$
in the ground state manifold, as they are of opposite parities;
and, 2)
transitions to high lying states with the same parities
are also suppressed due to the increased energy gaps.

We now estimate the adiabaticity requirement.
The parametric motion of $d[t]$ leads to a time dependent Hamiltonian
$H(t)$ with eigen-energy $E_n^{(p)}(t)$ and state
$\psi_n^{(p)}(z,d[t])$. Nonadiabatic transitions can be neglected only if
\begin{eqnarray}
\left|{{\langle \psi_i^{(p)}(z,d[t])|{d\over dt}|\psi_j^{(p')}(z,d[t])\rangle}\over {
E_i^{(p)}(t)-E_j^{(p')}(t)}}\right|\ll 1.
\label{ad}
\end{eqnarray}
As was pointed out in Ref. \cite{hansel0102}, nonzero matrix
elements can only exist among states with same parity because
$d/dt$ does not change parity. Further, because ${\langle
\psi_i^{(p)}(z,d[t])|{d/dt}|\psi_j^{(p')}(z,d[t])\rangle}
=|{{\langle
\psi_i^{(p)}(z,d[t])|[{dH(t)/dt}]|\psi_j^{(p')}(z,d[t])\rangle}|/
|{E_i^{(p)}(t)-E_j^{(p')}(t)}}|$, Eq. (\ref{ad}) can be easily
estimated. For eigenstates with fixed parity $p$, we find $\langle
\psi_i^{(p)}(z,d[t])|{d/dt}|\psi_i^{(p)}(z,d[t])\rangle=0$ as
$\psi_i^{(p)}(z,d[t])$ can always be chosen as a real function.

When only one trap moves as illustrated in Fig. \ref{fig1},
non-inertial effects of the center of mass frame have to be
accounted for.
Using $a(t)$ to denote the acceleration of the center
of mass frame, we can investigate the subsequent nonadiabatic effects with,
\begin{eqnarray}
i\hbar{\partial\over \partial t}\psi_n^{(p)}(z,d)=\left[{p_z^2\over 2M}+V_2(z,d)+Maz\right]
\psi_n^{(p)}(z,d).\hskip 6pt
\label{td}
\end{eqnarray}
The non-inertial force term $Maz$ breaks parity conservation, thus
will cause non-adiabatic transitions between different parity
states. This can seriously affect the splitting as the two
opposite parity states within the ground state family are always
closely spaced in the large $|d|$ limit. We have performed
detailed studies on the asymmetries between the two traps and on
the effect of fluctuations of $d(t)$. The effect of asymmetries
are found to influence the splitting only in the second order
\cite{zhang}. The tolerable levels for imperfections, {\it e.g.},
fluctuations and asymmetries, seem to be well within the current
experimental capabilities. These results as well as an interesting
extension of our scheme to more atoms and atomic condensates will
be presented elsewhere \cite{bohn}.

The quality performance of our splitting scheme resembles the
earlier adiabatic passage protocol with
a pair of counter-propagating Raman pulses in a temporally
counter-intuitive order \cite{zoller,weiss,aspect1}.
The Raman protocol involves only one adiabatic state,
or the famous dark state \cite{zoller},
thus it is only a beam splitter, not an interferometer as only
one adiabatic path is involved.
Our protocol on the other hand contains two interfering parts:
the symmetric and anti-symmetric states, thus does form a
complete interferometer. However, it is not
an universal interferometer because the amplitudes for the
two paths are
always equal to $1/\sqrt{2}$ \cite{ketterle2}, and the
interference is completely controlled by their relative phase.
In both cases, the restrictions from being an universal
interferometer give rise to higher fidelities provided
strict adiabaticity is maintained.

In conclusion, we have studied the splitting of an trapped atomic
wave-packet by moving one trap with respect to a second trap.
Through a careful analysis of the dependence of the single atom
energy levels on the trap separation, we have shown that our
splitting scheme makes use of a ``topologically invariant" like
quantity. Thus, our suggested approach is capable of demonstrating
a high level of fidelity against imprecise or imperfect control of
external parameters. To our knowledge, our results have not been
previously seen in other studies of optical or atomic
splitting/guiding systems. In view of the active experimental
efforts in this area \cite{kreutzmann04,aspect,natphys,sa}, our
scheme shines new light on atomic splitting with a time dependent
trap.

We thank Mr. H.H. Jen for
helpful confirmations of the numerical simulations.
L. Y. acknowledges Prof. W. Ketterle and Prof. A. Aspect for
insightful discussions.
This work is supported by NSF and NASA. M.Z, P.Z., and L.Y. also
acknowledge the support of NSFC.

\end{document}